\documentclass[12pt]{article}
\usepackage{natbib}
\usepackage{times}
\usepackage{graphicx}
\usepackage{apjfonts}
\usepackage{geometry}
\usepackage[utf8]{inputenc}
\usepackage{enumitem,amssymb}
\usepackage{ragged2e}
\newlist{thematic}{itemize}{8}
\setlist[thematic]{label=$\square$}
\usepackage{pifont}
\newcommand{\cmark}{\ding{51}}%
\newcommand{\done}{\rlap{$\square$}{\raisebox{2pt}{\large\hspace{1pt}\cmark}}%
\hspace{-2.5pt}}

\newcommand{\ltsim} {\lta} \newcommand{\gtsim} {\gta}
\newcommand{\msun}     {\ensuremath{{M}_{\scriptscriptstyle \odot}}}
\newcommand{\kms}      {\ensuremath{~\mathrm{km~s^{-1}}}}
\newcommand{\au}       {\ensuremath{~\mathrm{AU}}}
\newcommand{\ergs}     {\ensuremath{~\mathrm{erg\,s^{-1}}}}
\newcommand{\mhz}      {\ensuremath{~\mathrm{mHz}}}
\newcommand{\hz}       {\ensuremath{~\mathrm{Hz}}}
\newcommand{\yr}       {\ensuremath{~\mathrm{yr}}}
\newcommand{\kpc}      {\ensuremath{~\mathrm{kpc}}}
\newcommand{\Mpc}      {\ensuremath{~\mathrm{Mpc}}}
\newcommand{\msigma}   {\ensuremath{M}{--}\ensuremath{\sigma}}
\newcommand{\ml}       {\ensuremath{M}{--}\ensuremath{L}}
\newcommand{\mbh}      {\ensuremath{M_{\mathrm{BH}}}}
\newcommand{\apj}       {ApJ}
\newcommand{\nat}       {Nature}
\newcommand{\aap}       {A\&A}
\newcommand{\aj}        {AJ}
\newcommand{\mnras}     {MNRAS}
\newcommand{\apjs}      {ApJS}
\newcommand{\apjl}      {ApJ Letters}
\newcommand{\araa}      {Ann.\ Rev.\ Astronomy \& Astrophysics}
\newcommand{\prd}       {Phys.\ Rev.\ D}
\newcommand{\aapr}      {A\&A Review} 

\def\arcsec{\hbox{\ensuremath{^{\prime\prime}}}}

\def\farcs{\hbox{\ensuremath{.\!\!^{\prime\prime}}}}

\begin{document}
\raggedright
\huge
Astro2020 Science White Paper \linebreak

The Local Relics of of Supermassive Black Hole Seeds \linebreak
\normalsize

\noindent \textbf{Thematic Areas:} \hspace*{60pt} $\square$ Planetary Systems \hspace*{10pt} $\square$ Star and Planet Formation \hspace*{20pt}\linebreak
$\square$ Formation and Evolution of Compact Objects \hspace*{31pt} $\square$ Cosmology and Fundamental Physics \linebreak
  $\square$  Stars and Stellar Evolution~ \hspace*{1pt} \done\  Resolved Stellar Populations and their Environments \hspace*{40pt} \linebreak
  \done\   Galaxy Evolution   \hspace*{45pt} $\square$             Multi-Messenger Astronomy and Astrophysics \hspace*{65pt} \linebreak
  
\textbf{Principal Author:}

Name:	Jenny E Greene
 \linebreak						
Institution:  Princeton University
 \linebreak
Email: jgreene@astro.princeton.edu
 \linebreak
Phone:  914-924-9997
 \linebreak
 
\textbf{Co-authors:} (names and institutions)
  \linebreak
Aaron Barth (UC Irvine); 
Andrea Bellini (STScI);
Jillian Bellovary (Queensborough Community College/AMNH)
Kelly Holley-Bockelmann (Vanderbilt and Fisk);
Tuan Do (UCLA);
Elena Gallo (U of Michigan);
Karl Gebhardt (UT Austin);
Kayhan Gultekin (U Michigan);
Zoltan Haiman (Columbia University);
Matthew Hosek Jr. (UCLA);
Dongwon Kim (UC Berkeley);
Mattia Libralato (STScI);
Jessica Lu (UC Berkeley);
Kristina Nyland (National Research Council, resident at the Naval Research Laboratory);
Matthew Malkan (UCLA);
Amy Reines (Montana State U);
Anil Seth (U of Utah);
Tommaso Treu (UCLA);
Jonelle Walsh (Texas A \& M);
Joan Wrobel (NRAO)

\medskip

\justify

\textbf{Abstract:}
We have compelling evidence for \emph{stellar-mass} black holes (BHs) of $\sim 5-80$~\msun\ that form through the death of massive stars. We also have compelling evidence for so-called supermassive BHs ($10^5-10^{10}$~\msun) that are predominantly found in the centers of galaxies. 
We have very good reason to believe there must be BHs with masses in the gap between these ranges: the first $\sim 10^9$~\msun\ BHs are observed only hundreds of millions of years after the Big Bang, and all theoretically viable paths to making supermassive BHs require a stage of ``intermediate'' mass. However, {\it no BHs have yet been reliably detected in the $\textit{100}-\textit{10}^5$~\msun\ mass range}. Uncovering these intermediate-mass BHs is within reach in the coming decade. In this white paper we highlight the crucial role that 30-m class telescopes will play in dynamically detecting intermediate-mass black holes, should they exist. 

\pagebreak

\section{A Full Census of Black Holes in the Coming Decade}

The past ten years has transformed our understanding of black hole (BH) populations. The spectacular and surprising detection of gravitational waves from the merging of $\sim 30$~\msun\ black holes \citep{abbottetal2017} proved that we do not yet understand the full mass BH spectrum. At the same time, our quest to discover the first supermassive BHs has uncovered targets with \mbh$>10^9$~\msun\ at $z>7$ \citep[e.g.,][]{banadosetal2018}, underlining that the early history of these supermassive BHs is yet unknown. Meanwhile, time-domain surveys are uncovering diverse explosive events, which may be powered from stellar disruption by a heretofore unobserved class of $10^3-10^5$~\msun\ BHs \citep[e.g.,][]{linetal2018,perleyetal2019}. These developments set the stage to answer fundamental questions about the origin, growth, and demographics of the BH population:
\begin{itemize}
    \item How did supermassive BHs form?
    \item What and where are the seed BHs that are left over from this process?
    \item What is the full mass spectrum of BHs, particularly in the uncharted region between 100 and $10^5$~\msun?
\end{itemize}
In the coming decade, the advent of 30 meter-class, extremely large telescopes (ELTs) will enable direct dynamical searches for $10^3-10^5$~\msun\ BHs in our Local Volume. The mass distribution of such BHs will distinguish different seeding mechanisms \citep[e.g.,][]{ricartenatarajan2018}, is required to interpret detections of tidal disruption events with surveys like the Large Synoptic Survey Telescope \citep[LSST; e.g.,][]{stonemetzger2016}, and will inform searches for gravitational wave events \citep[e.g.,][]{fragioneetal2018}. Only ELTs will be able to dynamically measure BH masses in this unexplored intermediate-mass regime.


\section{Open Questions}

\subsection{Formation of the first massive black holes}

It remains a serious theoretical challenge to grow the first $10^9$~\msun\ BHs only hundreds of millions of years after the Big Bang. If the seeds of these BHs were formed from stellar processes, then even the death of the first stars likely cannot make a BH more massive than $\sim 100$~\msun\ \citep[e.g.,][]{madaurees2001}. Seeds formed as stellar remnants would be common. However, to reach $10^9$~\msun\ in such a short time, such a seed BH would have to grow at or above the Eddington limit continuously \citep[e.g.,][]{haimanloeb2001}. The timing challenge is significantly alleviated if the seeds are made heavy ($10^3-10^5$~\msun) through direct collapse of gas clouds, a very massive star phase, or dynamical runaway in a dense cluster \citep[see][for a review]{volonteri2010}. The consensus is that heavy seeds require special conditions and are expected to be rare \citep[but see ][]{dunnetal2018}. In contrast to massive, bulge-dominated galaxies containing high-mass BHs, low-mass galaxies provide a relatively pristine dynamical environment that preserves some memory of the seeding process \citep{bellovaryetal2019}. The overall occupation fraction of BHs in low-mass galaxies should be sensitive to different seeding mechanisms \citep{ricartenatarajan2018,sesana2011,haimanetal2019}.  Nascent cosmological simulations have yet to implement a fully self-consistent non-equilibrium chemical network that includes dust formation and radiative transfer over a wide energy range, let alone magnetohydrodynamics or the dynamical effects of gravitational wave recoil that may remove most BHs from globular clusters \citep[e.g.,][]{holley-bockelmannetal2008,fragioneetal2018}, so we still rely on these next-generation observations to constrain BH seed formation. 



\begin{figure}
\begin{center}
\includegraphics[width=6cm]{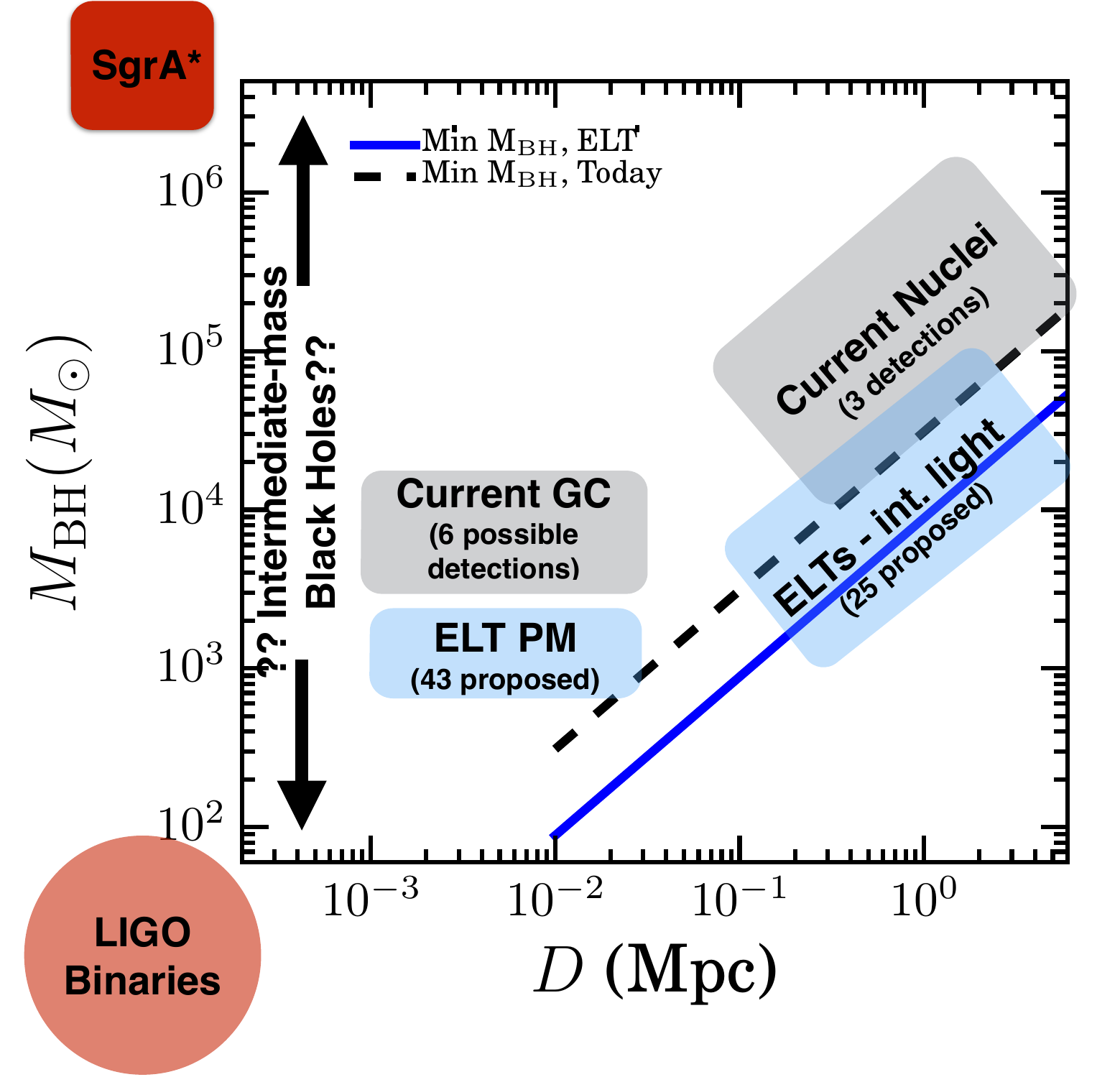}
\includegraphics[width=6cm]{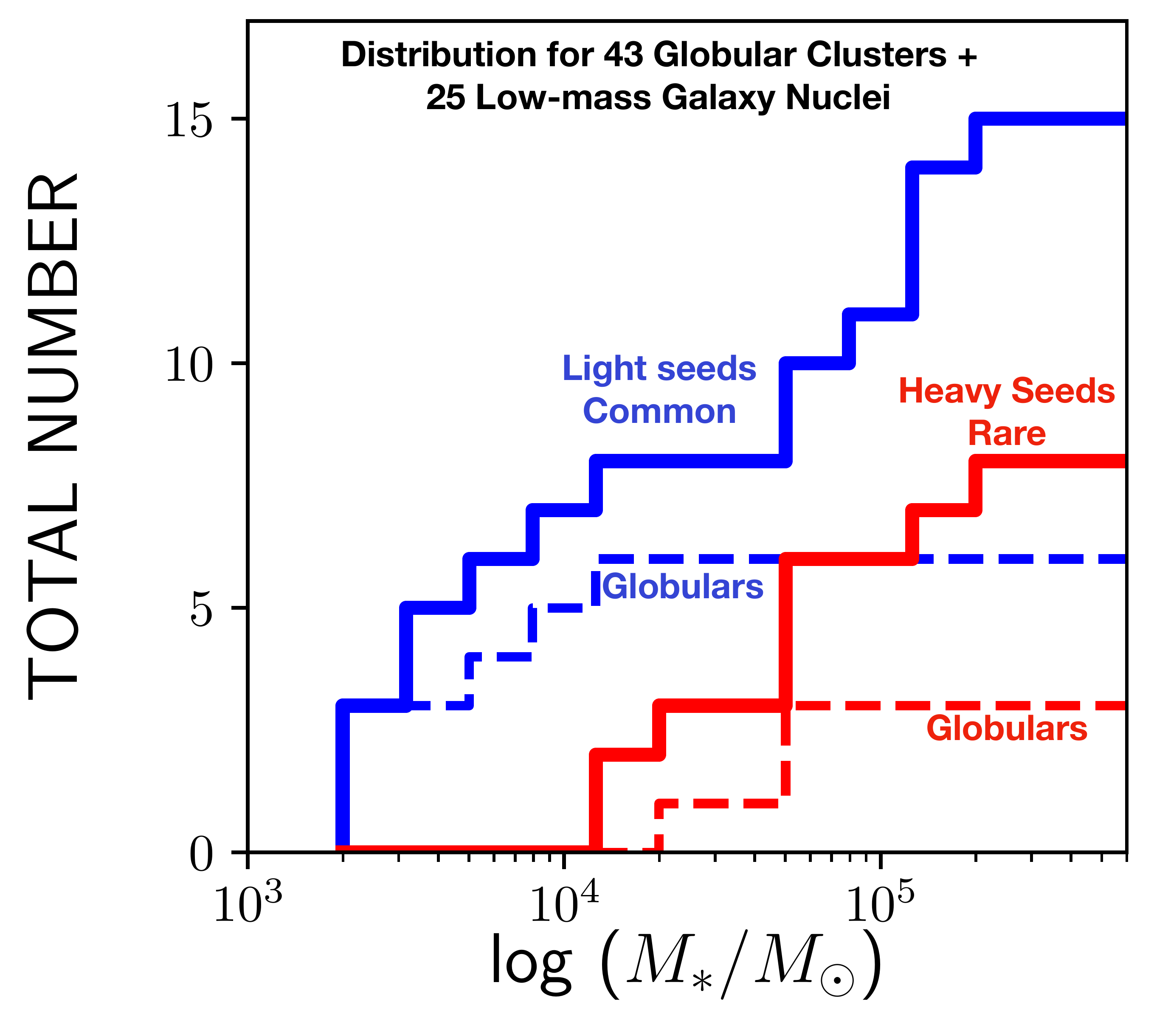}
\caption{
\small
{\it Left}: The landscape of dynamical IMBH detections within 5 Mpc.  We show the limiting BH mass with an ELT (blue solid line) assuming a $\sigma^* = 20$~\kms\ and a resolution limit of 0.01 arcsec as compared with current limits to 0.07 arcsec (black dashed). We schematically show the regions where existing limits are found. ELTs may provide detections nearly an order of magnitude better, depending on the unknown relationship between galaxy structure and \mbh. 
{\it Right}: The cumulative distribution of \mbh\ assuming a fiducial sample of 43 globular clusters and 25 galaxy nuclei, under two seeding scenarios (contribution from just the globulars shown in dashed). In the ``heavy'' seeding case, all the black holes have \mbh$=10^{4}-10^5$~\msun, the occupation fraction is 50\%, and very few clusters harbor seed BHs. In contrast, in light-seeding scenarios, most clusters harbor detectable BHs. If recoils remove BHs from globular clusters, then these numbers are too optimistic \citep[e.g.,][]{holley-bockelmannetal2008}.}
\label{f2}
\end{center}
\end{figure}

\vskip -10mm

\subsection{The search for \boldmath$<10^5$ M$_\odot$ BHs}

Quite a lot of progress has been made over the past decade to characterize the population of $\sim$10$^5$~\msun\ BHs in nearby galaxy nuclei.  Thanks to heroic efforts with integral-field units (IFUs) behind adaptive optics, there are now a handful of dynamical detections \citep{sethetal2010,denbroketal2015,nguyenetal2018} and limits \citep{gebhardtetal2001,vallurietal2004,barthetal2009,nguyenetal2018}. In addition, we see accretion signatures from a very small fraction of dwarf nuclei \citep[e.g.,][]{greeneho2004,milleretal2012,reinesetal2013,moranetal2014,pardoetal2016} and BH masses inferred for these using the dynamics of the broad-line region suggest systems $<10^5$~\msun\ in a few cases \citep{baldassareetal2015}. There are also dynamical detections of BHs in massive stellar clusters that likely started out as the nuclei of dwarf galaxies, but then were stripped by falling into the Local Group. The BH masses in these systems range from $10^4-10^7$ \msun\ \citep[][]{gebhardtetal2005,sethetal2014}; at the low mass end these detections remain very controversial \citep{vandermarelanderson2010}. In Galactic globular clusters, we see little evidence for massive BHs, with limits from accretion suggesting that there are few massive BHs $\gtrsim 1000$~\msun\ \citep{tremouetal2018} and dynamical observations mainly yielding upper limits. Ultra-Luminous X-ray sources also in principle may be powered by intermediate-mass BHs \citep{kaaretetal2017}. The most promising candidate to date is HLX-1 in the galaxy ESO243-49 \citep[e.g.,][]{farrelletal2009}, with an $L_X=10^{42}$ erg~s$^{-1}$ X-ray source showing variability and an X-ray spectrum suggestive of a BH \citep[e.g.,][]{soriaetal2017}.


\medskip
\noindent
{\it ELTs will provide order-of-magnitude improvements in the angular resolution that can be achieved for dynamical measurements. The ELTs will discover IMBHs with $10^3-10^5$~\msun\ should they exist.}


\section{Intermediate-mass Black Holes in the 2020's}

We anticipate a large range of breakthroughs in the 2020's in the realm of IMBH searches. In this white paper, we focus on the dramatic gains that can be made with volume-limited dynamical studies by exploiting the factor of $\sim 5$ improvement in angular resolution brought by the ELTs to resolve the gravitational sphere of influence of putative IMBHs ($\sim0.01$ arcsec for a $\sim 10^4$~\msun\ BH at 5 Mpc; Fig.\ 1).

\subsection{Using proper motions to find \boldmath$10^3-10^4$~M$_\odot$ BHs} 

\begin{figure}
\begin{center}
\includegraphics[width=0.9\textwidth]{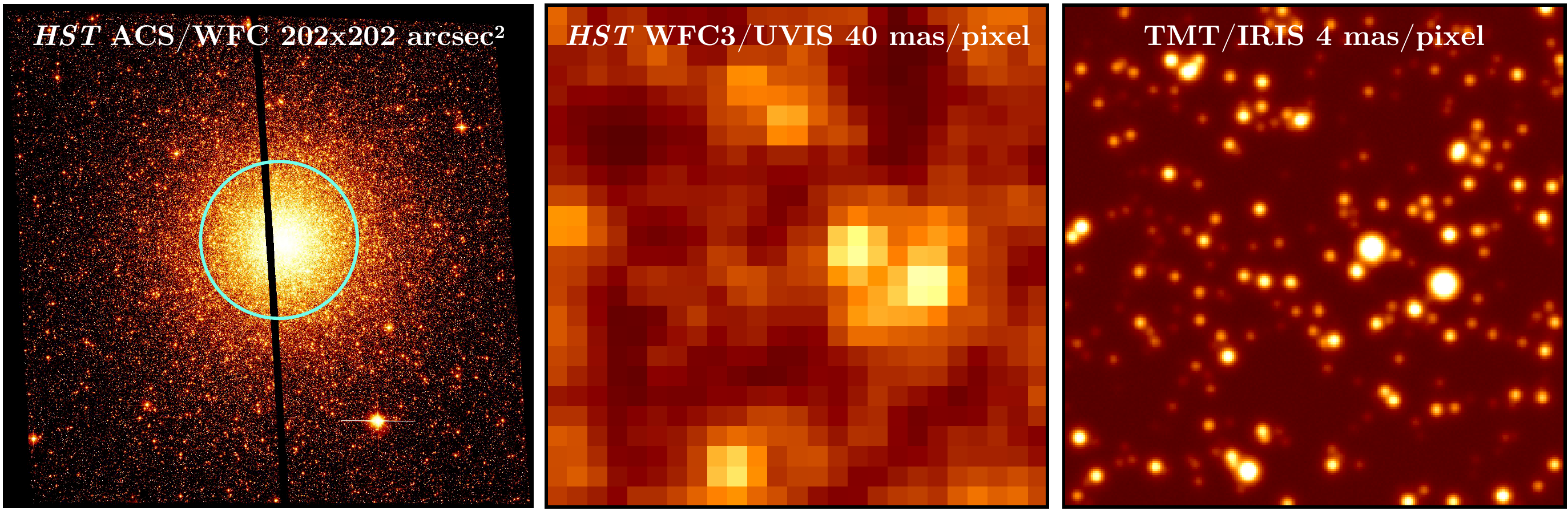}
\caption{
\small
{\it Left}: The GC NGC 6441 as seen by the \textit{HST}'s ACS/WFC. The total FoV is $202\times202$ arcsec$^2$, 50 mas/pixel. The blue circle marks the cluster's half-light radius ($34\farcs2$). {\it Middle}: The central 1 arcsec$^2$ as seen by the \textit{HST}'s WFC3/UVIS, the highest-resolution detector currently available on-board the \textit{HST} (40 mas/pixel; Note that JWST's NIRCAM will have similar resolution capabilities, with a pixel scale of 35 mas/pixel.) {\it Right}: Simulation of the same central 1 arcsec$^2$ as seen by TMT/IRIS.}
\label{f:hsttmt}
\end{center}
\end{figure}

High-precision proper-motion (PM) measurements with ELTs will enable groundbreaking improvements in sensitivity to infer the possible presence of IMBHs in globular clusters.  Proper motions require 3 micro-arcsec per year precision (\S4), but :\ (1) They can probe to fainter stars than spectroscopic studies, yielding better statistics on the kinematic quantities of interest;\ (2) Stars are measured individually, by contrast to integrated-light measurements, so a disproportionate contribution from bright giants is avoided;\ (3) Two components of velocity are measured, breaking the well-known degeneracy between increased mass and orbital anisotropy \citep{binneymamon1982}.

\medskip
\noindent
The crucial region needed to directly place constraints on the mass of a central BH corresponds to just the innermost few arcseconds of a cluster core. A BH causes a central increase in the velocity dispersion with $\sigma_v\propto R^{-1/2}$. Of course, a central (dark) mass concentration could be either a BH or the natural result of mass segregation.  By comparing the measured stellar motions with the most detailed available Fokker-Planck and N-body models it will be possible to measure (or provide upper limits for) the mass of any central IMBH \citep[e.g.,][]{lutzgendorfetal2013}.  The detection of any unusually fast-moving star (the so-called ``smoking gun'') would provide additional constraints on the presence of an IMBH. There are $\sim 40$ GCs more massive than $\sim 2 \times 10^5$~\msun\ in the Milky Way. With just about 100 hours of ELT observations taken over four years, it would be possible to obtain 3 epochs of PM data for these clusters (about 2.5 hours each), and constrain a possible central BH to $<2 \times 10^3$~\msun. Such data would constrain not only the presence of an IMBH (even wandering within the cluster core), but also the structure, dynamics, and evolution of GCs at high central densities in general \citep[e.g.,][]{libralatoetal2018}. 

\subsection{Integrated-light observations of \boldmath$10^4-10^5$~M$_\odot$ BHs in galaxy nuclei} 

ELTs equipped with adaptive optics and IFU spectrographs will dramatically improve the spatial resolution available to resolve the gravitational sphere of influence of BHs in the centers of galaxies, and the sensitivity gains will be transformative for detection of low-mass BHs \citep{doetal2014}. \emph{JWST} will not have the spectral nor spatial resolution needed to work on this problem. As shown in Figure 1, these measurements will be sensitive to the dynamical signatures of $\sim 10^3$~\msun\ BHs at the distance of Andromeda, and $10^4$~\msun\ BHs out to 5 Mpc, although the exact limits are uncertain because they depend on how \mbh\ scales with $\sigma^*$ \citep{lutzgendorfetal2013,greeneetal2016}. 


\medskip
\noindent
We envision a number of high-priority samples for IMBH searches. Within a volume of 5 Mpc, where a mass limit of $10^4$~\msun\ is possible for a wide range of $\sigma^*$, it will be possible to target all low-mass galaxies with $10^9<M^*/M_{\odot}<3\times 10^9$. It will also be fruitful to target all massive globular clusters in Andromeda \citep[like G1; e.g.,][]{gebhardtetal2005}. While some upper limits would ensue, we would also advocate targeting all galaxies down to a similar mass limit in Virgo \citep[e.g.,][]{coteetal2006} and/or Fornax \citep[e.g.,][]{ordenes-briceetal2018}. As a proof of concept, we consider the $\sim 25$ nucleated spiral galaxies within 5 Mpc, which would make an excellent first sample. In $\sim 200$ hours, it would be possible to obtain BH detections down to $\sim 10^4$~\msun\ for this full sample. When combined with the globular cluster study outlined above, we anticipate gaining significant new insights into whether seed BHs formed `heavy' or `light' (see Figure 1, right). Non-nucleated and lower-mass dwarf galaxies in the Local Group may be feasible targets for proper motion studies as described above, but will require better determinations of their dynamical centers before they can be the target of IFU observations.

\subsection{Synergies with other next-generation surveys}

There are many exciting prospects in the next decade that will complement (and provide targets for) ELT efforts. Gravitational wave searches with Advanced LIGO may constrain the early phases of dynamical run-away in clusters \citep{kovetzetal2018}, while satellites like LISA should detect the mergers of IMBHs should they occur \citep[e.g.,][]{sesanaetal2007,natarajanetal2019}, even at high redshift. LISA would also detect high mass ratio inspirals of IMBHs and stellar mass black holes~\citep{amaro-seoaneetal2017}. Predictions for LISA hinge on a better understanding of the IMBH mass function and their occupation fraction in dwarfs \citep[e.g.,][]{micic2011,bellovaryetal2019}. Multi-wavelength searches for accretion (including tidal disruption events) in low-mass galaxies with next-generation X-ray and radio telescopes \citep{galloetal2019,wrobeletal2019} will provide complementary constraints on occupation fractions \citep[e.g.,][]{milleretal2015}, as well as potentially promising targets for follow-up dynamical measurements. However, only the dynamical approaches considered here can measure the mass function of IMBHs in a volume-limited sample to compare with seeding models \citep[e.g.,][]{volonteri2010}.



\section{Recommendations}

While the current generation of 8-10m class telescopes equipped with adaptive optics has enabled great progress in exploring the demographics of supermassive BHs, a major leap in observational capabilities is required in order to push dynamical BH searches into the IMBH regime. The planned next generation of ELTs will have the revolutionary capability to determine whether there are BHs in the mass range $10^3-10^5$~\msun, and to determine their occupation fraction and mass distribution in nearby star clusters and galactic nuclei. The key requirements are adaptive optics imaging with high astrometric precision, and diffraction-limited integral-field spectroscopy with sufficient spectral resolution to map the internal kinematics of low-mass galactic nuclei and extragalactic star clusters. The critical new capabilities are summarized below.


\smallskip

\noindent
$\bullet~$ For astrometric IMBH searches based on proper motions, we require an angular resolution of better than 10 mas (or about $\sim 4$ mas/pixel with Nyquist sampling) to resolve stars all the way to the center of massive globular clusters (see, e.g., Fig.~\ref{f:hsttmt}).  Such angular resolution will be easily achievable by adaptive-optics imagers on 30m-class telescopes. In terms of astrometric precision, a $10^3$~\msun\ IMBH in a globular cluster produces a 1 km\,s$^{-1}$ increase in the central velocity dispersion. PM errors should not be larger than half the intrinsic velocity dispersion. A 3$\sigma$ detection implies PMs with errors $\lesssim$ 150 m\,s$^{-1}$. For a typical globular cluster distance of 10 kpc, this can be achieved with 3 epochs taken every two years (for a 4 year baseline) and 12 exposures per epoch (assuming a conservative single-exposure positional precision of 0.02 pixel). The gravitational influence radius of IMBHs in globular clusters is just a few arcsec, well within the typical FoV of AO imagers of about 0.5–1 arcmin.

\smallskip

\noindent
$\bullet~$ For integral-field spectroscopic observations of the 2.29 $\mu$m CO bandhead and the Ca~II near-IR triplet (8500 \AA), the optimal pixel scale would correspond to Nyquist sampling of the diffraction limit on an ELT over a field of view of $\sim0.5$ arcsecond or greater. Spectral resolving power should be at least $R=8000$ to measure kinematics in nuclear star clusters with velocity dispersions down to $15-20$ km s$^{-1}$, and $R=10,000$ will be required for the lowest-mass targets. An adaptive optics capability operating at the Ca~II triplet would enable resolution of the dynamical sphere of influence for \mbh\ as low as $\sim10^4$ \msun\ in galaxies out to 5 Mpc.


\smallskip

\noindent
$\bullet~$ To access the full sample of Milky Way and M31 globular clusters, Local Group dwarfs (such as the M31 satellites NGC 205 and NGC 185), and spiral galaxies within 5 Mpc, telescopes located in both the Northern and Southern hemispheres will be required.

\smallskip 

\noindent
$\bullet~$ Finally, we recommend project-level support for development of open, shared-use software for the dynamical modeling of the data proposed here. Just as data-reduction software is critical to science efficiency, so too is publicly available analysis code.


\pagebreak



\end{document}